\documentclass[pre,aps,twocolumn,showpacs,floatfix]{revtex4}
\usepackage[dvips]{graphicx}
\usepackage{amsmath}
\usepackage{amssymb}
\usepackage[nice]{nicefrac}
\usepackage{color}

\begin{document}

\title{Paradoxical diffusion: Discriminating between normal and anomalous random walks}

\author{Bart{\l}omiej Dybiec}
\email{bartek@th.if.uj.edu.pl}
\affiliation{Marian Smoluchowski Institute of Physics, and Mark Kac Center for Complex Systems Research, Jagellonian University, ul. Reymonta 4, 30--059 Krak\'ow, Poland}

\author{Ewa Gudowska-Nowak}
\email{gudowska@th.if.uj.edu.pl}
\affiliation{Marian Smoluchowski Institute of Physics, and Mark Kac Center for Complex Systems Research, Jagellonian University, ul. Reymonta 4, 30--059 Krak\'ow, Poland}

\date{\today}
\begin{abstract}
Commonly, normal diffusive behavior is characterized by a linear dependence of the second central moment
on time, $\langle x^2(t) \rangle\propto t$,
while anomalous behavior is expected to show a different time dependence,
$ \langle x^2(t) \rangle \propto t^{\delta}$ with $\delta <1$ for subdiffusive
and $\delta >1$ for superdiffusive motions. Here we demonstrate that this kind of qualification, if applied straightforwardly, may be misleading: There are anomalous transport motions revealing perfectly ``normal'' diffusive character ($\langle x^2(t) \rangle\propto t$), yet being non-Markov and non-Gaussian in nature.
We use recently developed framework \cite[Phys. Rev. E \textbf{75}, 056702 (2007)]{magdziarz2007b} of Monte Carlo simulations which incorporates anomalous diffusion statistics in time and space and creates trajectories of such an extended random walk. For special choice of stability indices describing statistics of waiting times and jump lengths, the ensemble analysis of paradoxical diffusion is shown to hide temporal memory effects which can be properly detected only by examination of formal criteria of Markovianity (fulfillment of the Chapman-Kolmogorov equation).
\end{abstract}

\maketitle

%
%
\section{Introduction}

Usually various types of diffusion processes are classified by analysis of the spread of the distance traveled by a random walker. If the mean square displacement grows like $\langle [x-x(0)]^2 \rangle \propto t^\delta$ with $\delta<1$ the motion is called subdiffusive, in contrast to normal ($\delta=1$) or superdiffusive ($\delta>1$) situations. For a free Brownian particle moving in one dimension, a stochastic random force entering its equation of motion is assumed to be composed of a large number of independent identical pulses. If they posses a finite variance, then by virtue of the standard Central Limit Theorem (CLT) the distribution of their sum follows the Gaussian statistics. However, as it has been proved by L\'evy and Khintchine, the CLT can be generalized for independent, identically distributed (i.i.d) variables characterized by non-finite variance or even non-finite mean value. With a L\'evy forcing characterized by a stability index $\alpha<2$ independent increments of the particle position sum up yielding $\langle [x-x(0)]^2 \rangle \propto t^{2/\alpha}$ \cite{bouchaud1990}, see below. Such
enhanced, fast superdiffusive motion is observed in various real situations when a test particle is able to perform unusually large jumps \cite{dubkov2008,shlesinger1995,metzler2004}. L\'evy flights have been documented to describe motion of fluorescent probes in living polymers, tracer particles in rotating flows and cooled atoms in laser fields. They serve also as a paradigm of efficient searching strategies \cite{Viswanathan1996,sims2008,reynolds2009} in social and environmental problems with some level of controversy \cite{Edwards2007}.

In contrast, transport in porous, fractal-like media or relaxation kinetics in inhomogeneous materials are usually ultraslow, i.e. subdiffusive \cite{klafter1987,metzler2000,metzler2004}. The most intriguing situations take place however, when both effects -- occurrence of long jumps and long waiting times for the next step -- are incorporated in the same scenario \cite{zumofen1995}. The approach to this kind of anomalous motion is provided by continuous time random walks (CTRW) which assume that the steps of the walker occur at random times generated by a renewal process.
In particular, a mathematical idealization of a free Brownian motion (Wiener process $W(t)$) can be then derived as a limit (in distribution) of i.i.d random (Gaussian) jumps taken at infinitesimally short time intervals of non-random length. Other generalizations are also possible, e.g. $W(t)$ can be defined as a limit of random Gaussian jumps performed at random Poissonian times. The characteristic feature of the Gaussian Wiener process is the continuity of its sample paths. In other words, realizations (trajectories) of the Wiener process are continuous (although nowhere differentiable) \cite{doob1942}. The process is also self-similar (scale invariant) which means that by rescaling $t'=\lambda t$ and $W'(t)=\lambda^{-1/2}W(\lambda t)$ another Wiener process with the same properties is obtained. Among scale invariant stable processes, the Wiener process is the only one which possesses finite variance \cite{janicki1994,saichev1997,doob1942,dubkov2005b}. Moreover, since the correlation function of increments $\Delta W(s)= W(t+s)-W(t)$ depends only on time difference $s$ and increments of non-overlapping times are statistically independent, the formal differentiation of $W(t)$ yields a white, memoryless Gaussian process \cite{vankampen1981}:
\begin{equation}
\dot{W}(t)=\xi(t),\qquad \langle \xi(t)\xi(t') \rangle =\delta(t-t')
\end{equation}
commonly used as a source of idealized environmental noises within the Langevin description
\begin{equation}
dX(t)=f(X)dt + dW(t).
\end{equation}
Here $f(X)$ stands for the drift term which in the case of a one-dimensional overdamped motion is directly related to the potential $V(X)$, i.e. $f(X)= -dV(X)/dX$.

In more general terms the CTRW concept may asymptotically lead to non-Markov, space-time fractional noise $\tilde{\xi}(t)$, and in effect, to space-time fractional diffusion. For example, let us define $
\Delta \tilde{W}(t)\equiv \Delta X(t)=\sum^{N(t)}_{i=1}X_i,$
where the number of summands $N(t)$ is statistically independent from $X_i$ and governed by a renewal process $\sum^{N(t)}_{i=1}T_i\leqslant t < \sum^{N(t)+1}_{i=1}T_i$ with $t>0$.
Let us assume further that $T_i$, $X_i$ belong to the domain of attraction of stable distributions, $T_i\sim S_{\nu,1}$ and $X_i\sim S_{\alpha,\beta}$, whose corresponding characteristic functions $\phi(k)= \langle \exp(ikS_{\alpha,\beta}) \rangle=\int^{\infty}_{-\infty}e^{ikx} l_{\alpha,\beta}(x;\sigma) dx$, with the density $l_{\alpha,\beta}(x;\sigma)$, are given by
\begin{equation}
\phi(k) = \exp\left[ -\sigma^\alpha|k|^\alpha\left( 1-i\beta\mathrm{sign}k\tan
\frac{\pi\alpha}{2} \right) \right],
\label{eq:charakt1}
\end{equation}
for $\alpha\neq 1$
and
\begin{equation}
\phi(k) = \exp\left[ -\sigma|k|\left( 1+i\beta\frac{2}{\pi}\mathrm{sign}k\log|k| \right) \right].
\label{eq:charakt2}
\end{equation}
for $\alpha=1$.
Here the parameter $\alpha\in(0,2]$ denotes the stability index, yielding the asymptotic long tail power law for the $x$-distribution, which for $\alpha<2$ is of the $|x|^{-(1+\alpha)}$ type. The parameter $\sigma$ ($\sigma\in(0,\infty)$) characterizes the scale whereas $\beta$ ($\beta\in[-1,1]$) defines an asymmetry (skewness) of the distribution.

Note, that for $0<\nu<1$, $\beta=1$, the stable variable $S_{\nu,1}$ is defined on positive semi-axis. Within the above formulation the counting process ${N(t)}$ satisfies
\begin{eqnarray}
&& \lim_{t\rightarrow \infty}\mathrm{Prob}\left\{ \frac{N(t)}{(t/c)^{\nu}}<x\right\} =
 \lim_{t\rightarrow \infty}\mathrm{Prob}\left\{ \sum_{i=1}^{[(t/c)^{\nu}x]}T_i>t\right\}\nonumber \\
&& = \lim_{n\rightarrow\infty}\mathrm{Prob}\left\{ \sum_{i=1}^{[n]}T_i>\frac{cn^{1/\nu}}{x^{1/\nu}}\right\} \\
&& = \lim_{n\rightarrow\infty}\mathrm{Prob}\left\{ \frac{1}{cn^{1/\nu}}\sum_{i=1}^{[n]}T_i>\frac{1}{x^{1/\nu}}\right\}
 = 1-L_{\nu,1}(x^{-1/\nu}), \nonumber
\end{eqnarray}
where $[(t/c)^{\nu}x]$ denotes the integer part of the number $(t/c)^{\nu}x$ and $L_{\alpha,\beta}(x)$ stands for the stable distribution of random variable $S_{\alpha,\beta}$, i.e. $l_{\alpha,\beta}(x)=dL_{\alpha,\beta}(x)/dx$.
Moreover, since
\begin{equation}
\lim_{n\rightarrow\infty}\mathrm{Prob}\left\{ \frac{1}{c_1n^{1/\alpha}}\sum_{i=1}^{n}X_i<x\right\}\rightarrow L_{\alpha,\beta}(x)
\end{equation}
and
\begin{equation}
 p(x,t)=\sum_{n}p(x|n)p_n(n(t)),
 \label{eq:suma}
\end{equation}
asymptotically one gets
\begin{equation}
p(x,t)\sim (c_2t)^{-\nu/\alpha}\int_0^{\infty} l_{\alpha,\beta}\left((c_2t)^{-\nu/\alpha}x\tau^{\nu/\alpha}\right)l_{\nu,1}(\tau)\tau^{\nu/\alpha}d\tau,
\label{eq:solution}
\end{equation}
where $c_1$ and $c_2$ are constants.
The resulting (in general non-Markov) process becomes $\nu/\alpha$ self-similar L\'evy random walk \cite{tunaley1974,saichev1997,shlesinger1995,kotulski1995,nielsen2001,barkai2002,uchaikin2003,srokowski2009}, i.e.
\begin{equation}
 p(x,t)=t^{-\nu/\alpha}p(xt^{-\nu/\alpha},1).
\label{eq:scaling}
\end{equation}

The asymptotic form given by Eq.~(\ref{eq:solution}) can be easily derived \cite{scalas2006,saichev1997,barkai2002,germano2009} for decoupled CTRW by applying Tauberian theorems to the Montroll-Weiss \cite{montroll1965} expression
\begin{eqnarray}
p(q,u) & = & \int^{\infty}_0 dt\int^{+\infty}_{-\infty} dxe^{-ut+iqx}p(x,t)\nonumber \\
& = & \frac{1-\psi(u)}{u}\frac{1}{1-w(q)\psi(u)}
\label{eq:lap}
\end{eqnarray}
for the Laplace-Fourier transform of $p(x,t)$. The latter satisfies the integral (master) equation
\begin{eqnarray}
p(x,t) & = & \delta(x)\left [1-\int^t_0 \psi(t)dt\right ] \nonumber \\
& + & \int_0^t\psi(t-s)\left[\int^{\infty}_{-\infty}w(x-y)p(y,t)dy\right]ds.
\label{eq:master}
\end{eqnarray}
Here $\psi(t)$ stands for a waiting time probability density function (PDF) whereas $w(x)$ denotes a jump length PDF. With a suitable change of time and space variables, in the limit of $x\rightarrow\infty$, $t\rightarrow\infty$, the Laplace-Fourier transform $p(q,u)$ can be written as
\begin{eqnarray}
p(q,u) & = & \frac{u^{\nu-1}}{u^{\nu}+|q|^{\alpha}}\nonumber \\
& = & u^{\nu-1}\int_0^{\infty}ds \exp\left[-s(u^{\nu}+|q|^{\alpha})\right].
\end{eqnarray}
The inverse Laplace transform of $p(q,u)$ can be expressed in a series form \cite{saichev1997}:
\begin{eqnarray}
p(q,t) & = & \frac{1}{2\pi i}\int^{c+\infty}_{c-\infty}p(q,u)e^{ut}du\nonumber \\
& = & \sum^{\infty}_{k=0}\frac{(-1)^k}{\Gamma(k\nu+1)}(|q|^{\alpha}t^{\nu})^k\nonumber \\
& = & E_{\nu}\left(-|q|^{\alpha}t^{\nu}\right),
\end{eqnarray}
where $E_{\nu}(z)$ is the Mittag-Leffler function of order $\nu$
\begin{equation}
E_{\nu}(z)=\sum^{\infty}_{k=0}\frac{z^k}{\Gamma(k\nu+1)}.
\label{eq:mittag}
\end{equation}
Further application of the inverse Fourier transform in $q$ yields \cite{saichev1997} a final series representation of $p(x,t)$:
\begin{eqnarray}
p(x,t) & = & \frac{1}{\pi|y|t^{\nu/\alpha}}\sum^{\infty}_{k=0}\frac{(-1)^k}{|y|^{k\alpha}} \frac{\Gamma(k\alpha+1)}{\Gamma(k\nu+1)}\nonumber \\
& & \times   \cos\left[\frac{\pi}{2}(k\alpha+1)\right]
\label{eq:series}
\end{eqnarray}
where $y=x/t^{\nu/\alpha}$. The above series are divergent for $\alpha\geqslant\nu$. However, for $\alpha=\nu$, the summation has been shown \cite{saichev1997} to produce the closed analytical formula
\begin{equation}
p(x,t)=\frac{1}{\pi |y|t}\frac{\sin(\pi\nu/2)}{|y|^{\nu}+|y|^{-\nu}+2\cos(\pi\nu/2)}
\label{eq:closedformula}
\end{equation}
where (as previously) $y=x/t^{\nu / \alpha}$.

The exact solution of the decoupled CTRW, as given by the infinite sum~(\ref{eq:suma}) of stable probability densities, has been studied by Barkai \cite{barkai2002} based on a special choice of PDFs $w(x)$ and $\psi(t)$. In particular, in \cite{barkai2002} the extremely slow convergence of certain CTRW solutions to the (fractional) diffusion approximation has been discussed. The rigorous proof of equivalence between some classes of CTRWs and fractional diffusion has been given by Hilfer and Anton \cite{hilfer1995}

In this article we investigate CTRW scenarios which, in an asymptotic limit, yield paradoxical diffusion, i.e. the non-Markovian superdiffusive process taking place under sublinear operational time. The combination of long flights and long breaks between them is responsible for the characteristic shape of the PDF and scaling properties of moments. In particular, for $\alpha=2\nu$, the paradoxical diffusion process exhibits the same scaling as an ordinary Brownian motion despite its PDF is significantly different from Gaussian.

In a forthcoming Section, the relation between fractional calculus and CTRW approach is briefly reminded and the experimental results based on numerical PDF estimators are presented. Section III is devoted to the discussion of scaling properties of moments. In Section IV detection and analysis of memory effects in empirical series of the CTRW-type realizations are proposed and critically tested.

%
%
\section{Relation between CTRW and fractional calculus}

The theory of stochastic integration of a corresponding Ito-Langevin equation with respect to a general CTRW ``measure'' $d\tilde{W}(t)=dX$ has been developed in a series of papers \cite{scalas2006,meerschaert2006,magdziarz2007,germano2009}. Here, we
study statistical properties of such a motion constrained to the initial position $X(0)=0$. To achieve the goals, we adhere to the scheme of stochastic subordination \cite{magdziarz2007b,magdziarz2008,magdziarz2007}, i.e.
we obtain the process of primary interest $X(t)$ as a function $X(t)=\tilde{X}(S_t)$ by randomizing the time clock of the process $X(s)$ using a different clock $S_t$. In this approach $S_t$ stands for $\nu$-stable subordinator,
$S_t=\mathrm{inf}\left \{s:U(s)>t \right \}$, where $U(s)$ denotes a strictly increasing $\nu$-stable process whose distribution $L_{\nu,1}$ yields a Laplace transform $\langle e^{-kU(s)} \rangle =e^{-sk^{\nu}}$. The parent process $\tilde{X}(s)$ is composed of increments of symmetric $\alpha$-stable motion described in an operational time $s$
\begin{equation}
d\tilde{X}(s)=-V'(\tilde{X}(s))ds+dL_{\alpha,0}(s),
\label{eq:langevineq}
\end{equation}
and in every jump moment the relation $U(S_t)=t$ is fulfilled. The (inverse-time) subordinator $S_t$ is (in general) non-Markovian hence, as it will be shown, the diffusion process $\tilde{X}(S_t)$ possesses also some degree of memory. The above setup has been recently proved \cite{magdziarz2007b,magdziarz2008,magdziarz2007} to give a proper stochastic realization of the random process described otherwise by a fractional diffusion equation:
\begin{equation}
 \frac{\partial p(x,t)}{\partial t}={}_{0}D^{1-\nu}_{t}\left[ \frac{\partial}{\partial x} V'(x) + \frac{\partial^\alpha}{\partial |x|^\alpha} \right] p(x,t),
\label{eq:ffpe}
\end{equation}
with the initial condition $p(x,0)=\delta(x)$. In the above equation ${}_{0}D^{1-\nu}_{t}$ denotes the Riemannn-Liouville fractional derivative ${}_{0}D^{1-\nu}_{t}=\frac{\partial}{\partial t}{}_{0}D^{-\nu}_{t}$ defined by the relation
\begin{equation}
{}_{0}D^{1-\nu}_{t}f(x,t)=\frac{1}{\Gamma(\nu)}\frac{\partial}{\partial t}\int^{t}_0 dt'\frac{f(x,t')}{(t-t')^{1-\nu}}
\end{equation}
and $\frac{\partial^{\alpha}}{\partial |x|{\alpha}}$ stands for the Riesz fractional derivative with the Fourier transform ${\cal{F}}[\frac{\partial^{\alpha}}{\partial |x|^{\alpha}} f(x)]=-|k|^{\alpha}\hat{f}(x)$.
Eq.~(\ref{eq:ffpe}) has been otherwise derived from a generalized Master equation \cite{metzler1999}.
The formal solution to Eq.~(\ref{eq:ffpe}) can be written \cite{metzler1999} as:
\begin{equation}
p(x,t)=E_{\nu}\left (\left[ \frac{\partial}{\partial x} V'(x) + \frac{\partial^\alpha}{\partial |x|^\alpha} \right]t^{\nu}\right)p(x,0).
\end{equation}
For processes with survival function $\Psi(t)=1-\int_0^t\psi(\tau)d\tau$ (cf. Eq.~(\ref{eq:master})) given by the Mittag-Leffler function Eq.~(\ref{eq:mittag}), this solution takes an explicit form
\cite{saichev1997,jespersen1999,metzler1999,scalas2006,germano2009}
\begin{equation}
p(x,t)
=\sum_{n=0}^{\infty}\frac{t^{\nu n}}{n!}E^{(n)}_{\nu}(-t^{\nu})w_n(x)
\end{equation}
where $E^{(n)}_{\nu}(z)=\frac{d^n}{dz^n}E_{\nu}(z)$ and $w_n(x)\propto l_{\alpha,0}(x)$, see \cite{scalas2004,scalas2006}.

In this paper, instead of investigating properties of an analytical solution to Eq.~(\ref{eq:ffpe}), we switch to a Monte Carlo method \cite{magdziarz2007b,magdziarz2008,magdziarz2007,gorenflo2002,meerschaert2004} which allows generating trajectories of the subordinated process $X(t)$ with the parent process $\tilde{X}(s)$ in the potential free case, i.e. for $V(x)=0$. The assumed algorithm provides means to examine the competition between subdiffusion (controlled by a $\nu$-parameter) and L\'evy flights characterized by a stability index $\alpha$. From the ensemble of simulated trajectories the estimator of the density $p(x,t)$ is reconstructed and statistical qualifiers (like quantiles) are derived and analyzed.

\begin{figure}[!ht]
\begin{center}
\includegraphics[angle=0,width=8.0cm]{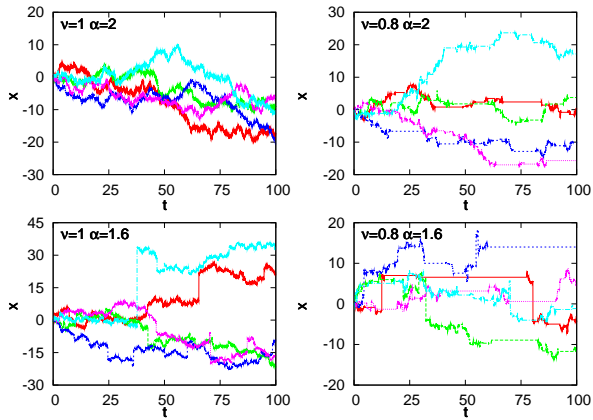}
\caption{Sample trajectories for $\nu=1, \alpha=2$ (left top panel), $\nu=1, \alpha=1.6$ (left bottom panel), $\nu=0.8, \alpha=2$ (right top panel) and $\nu=0.8, \alpha=1.6$ (right bottom panel). Eq.~(\ref{eq:langevineq}) was numerically approximated by subordination techniques with the time step of integration $\Delta t=10^{-2}$ and averaged over $N=10^6$ realizations.}
\label{fig:trajectories}
\end{center}
\end{figure}

As mentioned, the studied process is $\nu/\alpha$ self-similar (cf. Eq.~(\ref{eq:scaling})). We further focus on examination of a special case for which $\nu/\alpha=1/2$. As an exemplary values of model parameters we choose $\nu=1,\alpha=2$ (Markovian Brownian diffusion) and $\nu=0.8,\alpha=1.6$ (subordination of non-Markovian sub-diffusion with L\'evy flights). Additionally we use $\nu=1,\alpha=1.6$ and $\nu=0.8,\alpha=2$ as Markovian and non-Markovian counterparts of main cases analyzed. Fig.~\ref{fig:trajectories} compares trajectories for all exemplary values of $\nu$ and $\alpha$. Straight horizontal lines (for $\nu=0.8$) correspond to particle trapping while straight vertical lines (for $\alpha=1.6$) correspond to L\'evy flights. The straight lines manifest anomalous character of diffusive process.

\begin{figure}[!ht]
\begin{center}
\includegraphics[angle=0,width=8.0cm]{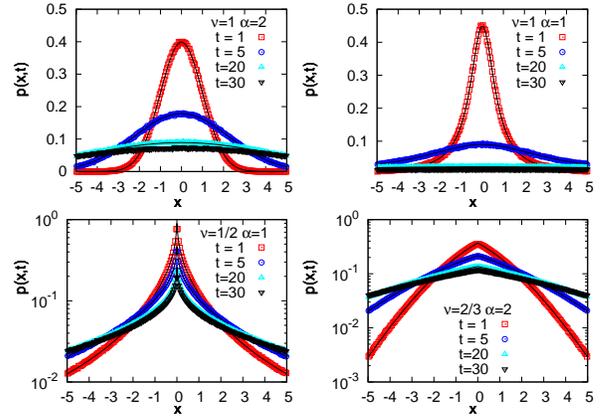}
\caption{(Color online) PDFs for $\nu=1$ with $\alpha=2$ (left top panel), $\nu=1$ with $\alpha=1$ (right top panel), $\nu=1/2$ with $\alpha=1$ (left bottom panel) and $\nu=2/3$ with $\alpha=2$ (right bottom panel). Eq.~(\ref{eq:langevineq}) was numerically approximated by subordination techniques with the time step of integration $\Delta t=10^{-3}$ and averaged over $N=10^6$ realizations. Solid lines present theoretical densities: Gaussian (left top panel), Cauchy (right top panel) and the $p(x,t)$ given by Eqs.~(\ref{eq:pa1n12}) (left bottom panel) and (\ref{eq:pa2n23}) (right bottom panel). Note the semi-log scale in the bottom panels.}
\label{fig:numeric-theory}
\end{center}
\end{figure}

To further verify correctness of the implemented version of the subordination algorithm \cite{{magdziarz2007b}}, we have performed extensive numerical tests. In particular, some of the estimated probability densities have been compared with their analytical representations and the perfect match between numerical data and analytical results have been found. Fig.~\ref{fig:numeric-theory} displays numerical estimators of PDFs and analytical results for $\nu=1$ with $\alpha=2$ (Gaussian case, left top panel), $\nu=1$ with $\alpha=1$ (Cauchy case, right top panel), $\nu=1/2$ with $\alpha=1$ (left bottom panel) and $\nu=2/3$ with $\alpha=2$ (right bottom panel). For those last two cases, the expressions for $p(x,t)$ has been derived \cite{saichev1997}, starting from the series representation given by Eq.~(\ref{eq:series}). For $\nu=1/2,\;\alpha=1$ the appropriate formula reads
\begin{equation}
p(x,t)=-\frac{1}{2\pi^{3/2}\sqrt{t}}\exp\left(\frac{x^2}{4t}\right)\mathrm{Ei}\left(-\frac{x^2}{4t}\right),
\label{eq:pa1n12}
\end{equation}
while for $\nu=2/3,\;\alpha=2$ the probability density is
\begin{equation}
p(x,t)=\frac{3^{2/3}}{2t^{1/3}}\mathrm{Ai}\left[ \frac{|x|}{(3t)^{1/3}} \right].
 \label{eq:pa2n23}
\end{equation}
$\mathrm{Ei}(x)$ and $\mathrm{Ai}(x)$ are the integral exponential function and the Airy function respectively.
We have also compared results of simulations and Eq.~(\ref{eq:closedformula}) for other sets of parameters $\nu$, $\alpha$. Also there, the excellent agreement has been detected (results not shown).

\begin{figure}[!ht]
\begin{center}
\includegraphics[angle=0,width=8.0cm]{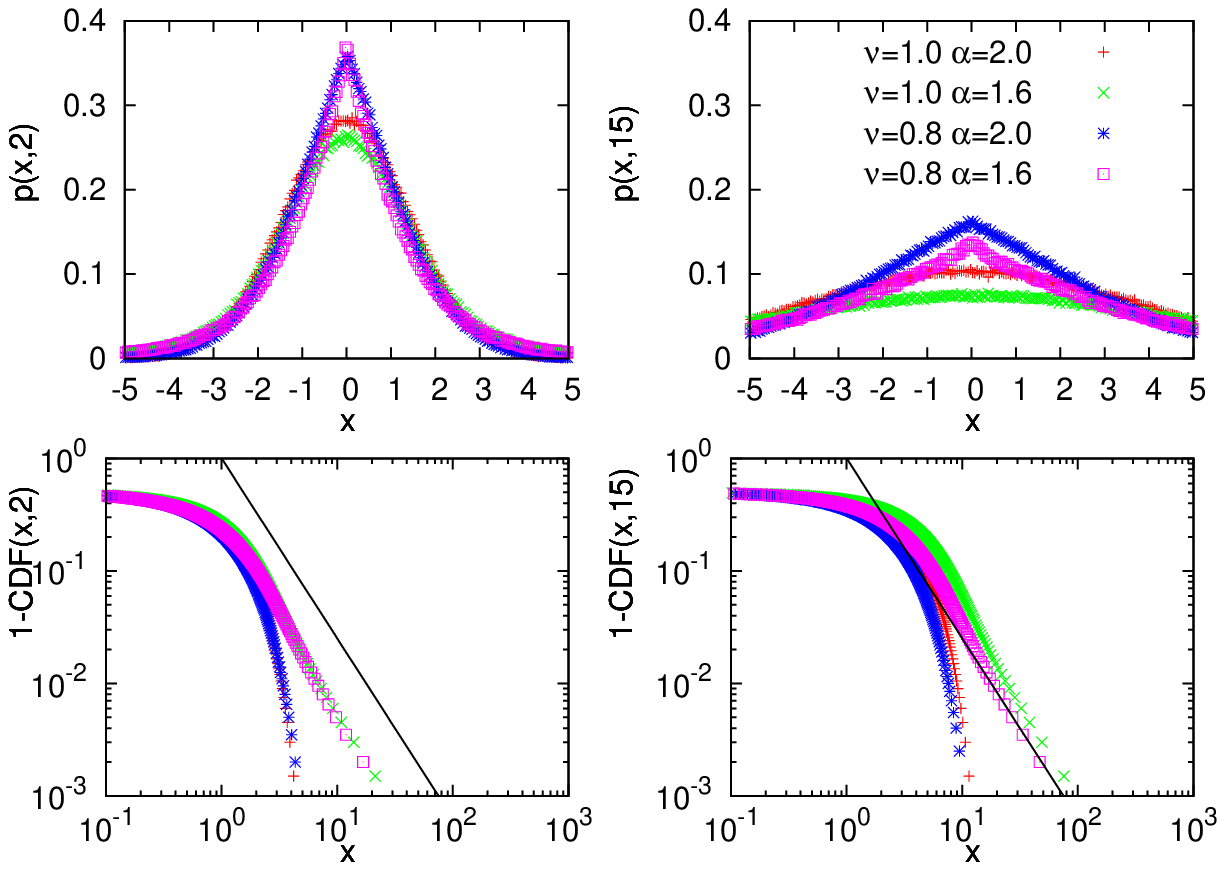}
\caption{(Color online) PDFs (top panel) and $1-\mathrm{CDF}(x,t)$ (bottom panel) at $t=2$ (left panel), $t=15$ (right panel). Simulation parameters as in Fig.~\ref{fig:trajectories}. Eq.~(\ref{eq:langevineq}) was numerically approximated by subordination techniques with the time step of integration $\Delta t=10^{-3}$ and averaged over $N=10^6$ realizations. Solid lines present theoretical asymptotic $x^{-1.6}$ scaling representative for $\alpha=1.6$ and $\nu=1$, i.e. for Markovian L\'evy flight.}
\label{fig:histograms}
\end{center}
\end{figure}

\begin{figure}[!ht]
\begin{center}
\includegraphics[angle=0,width=8.0cm]{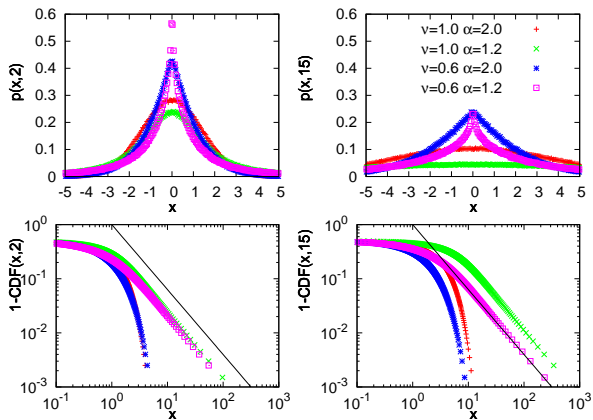}
\caption{(Color online) PDFs (top panel) and $1-\mathrm{CDF}(x,t)$ (bottom panel) at $t=2$ (left panel), $t=15$ (right panel). Simulation parameters as in Fig.~\ref{fig:trajectories}. Eq.~(\ref{eq:langevineq}) was numerically approximated by subordination techniques with the time step of integration $\Delta t=10^{-2}$ and averaged over $N=10^6$ realizations. Solid lines present theoretical asymptotic $x^{-1.2}$ scaling representative for $\alpha=1.2$ and $\nu=1$, i.e. for Markovian L\'evy flight.}
\label{fig:histograms12}
\end{center}
\end{figure}

Figure~\ref{fig:histograms} and \ref{fig:histograms12} display time-dependent probability densities $p(x,t)$ and corresponding cumulative distribution functions ($CDF(x,t)=\int_{-\infty}^xp(x',t)dx'$) for ``short'' and for, approximately, an order of magnitude ``longer'' times. The persistent cusp \cite{sokolov2002} located at $x=0$ is a finger-print of the initial condition $p(x,0)=\delta(x)$ and is typically recorded for subdiffusion induced by the subordinator $S_t$ with $\nu<1$. For Markov L\'evy-Wiener process \cite{dybiec2006,denisov2008} for which the characteristic exponent $\nu=1$, the cusp disappears and PDFs of the process $\tilde{X}(S_t)$ become smooth at $x=0$. In particular, for the Markovian Gaussian case ($\nu=1$, $\alpha=2$) corresponding to a standard Wiener diffusion, PDF perfectly coincides with the analytical normal density $N(0,\sqrt{t})$.

The presence of L\'evy flights is also well visible in the power-law asymptotic of CDF, see bottom panels of Figs.~\ref{fig:histograms} and \ref{fig:histograms12}. Indeed, for $\alpha<2$ independently of the actual value of the subdiffusion parameter $\nu$ and at arbitrary time, $p(x,t)\propto |x|^{-(\alpha+1)}$ for $x\rightarrow \infty$. Furthermore, all PDFs are symmetric with median and modal values located at the origin.

\begin{figure}[!ht]
\begin{center}
\includegraphics[angle=0,width=8.0cm]{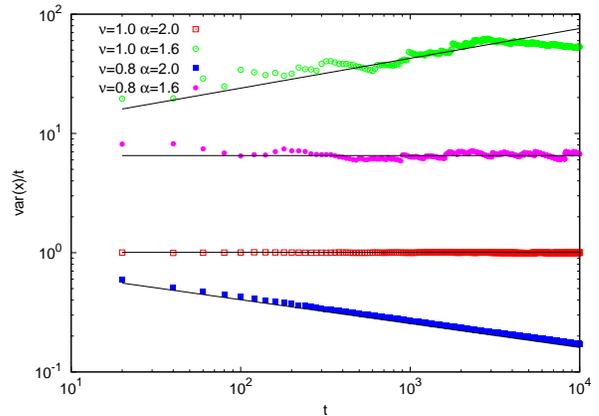}
\caption{(Color online) Time dependence of $\mathrm{var}(x)/t$. Straight lines present $t^{2\nu/\alpha-1}$ theoretical scaling (see Eq.~(\ref{eq:variancescaling}) and explanation in the text). Simulation parameters as in Fig.~\ref{fig:trajectories}.}
\label{fig:stdev}
\end{center}
\end{figure}

%
%
\section{Scaling properties of moments}

\begin{figure}[!ht]
\begin{center}
\includegraphics[angle=0,width=8.0cm]{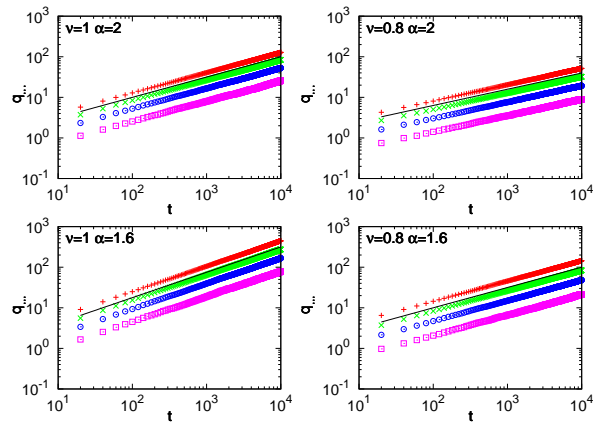}
\caption{(Color online) Quantiles: $q_{0.9}$, $q_{0.8}$, $q_{0.7}$, $q_{0.6}$ (from top to bottom) for $\nu=1, \alpha=2$ (left top panel), $\nu=1, \alpha=1.6$ (left bottom panel), $\nu=0.8, \alpha=2$ (right top panel) and $\nu=0.8, \alpha=1.6$ (right bottom panel). The straight line presents theoretical $t^{\nu/\alpha}$ scaling. Simulation parameters as in Fig.~\ref{fig:trajectories}.}
\label{fig:quantiles}
\end{center}
\end{figure}

The $\nu/\alpha$ self-similar character of the process (cf. Eq.~(\ref{eq:scaling})) is an outcome of allowed long flights and long breaks between successive steps. In consequence, the whole distribution scales as a function of $x/t^{\nu/\alpha}$ with the width of the distribution growing superdiffusively for $\alpha<2\nu$ and subdiffusively for $\alpha > 2\nu$. This $t^{\nu/\alpha}$ scaling is also clearly observable in the behavior of the standard deviation and quantiles $q_p(t)$, defined via the relation $\mathrm{Prob}\left \{X(t)\leqslant q_p(t)\right \}=p$, see Figs.~\ref{fig:stdev}, \ref{fig:quantiles}. For random walks subject to superdiffusive, long-ranging trajectories ($\alpha=1.6$), the asymptotic scaling is observed for sufficiently long times, cf. Fig.~\ref{fig:stdev}. On the other hand, normal (Gaussian) distribution of jumps superimposed on subdiffusive motion of trapped particles ($\nu =0.8$) clearly shows rapid convergence to the $\nu/\alpha$ law. Notably, both sets $\nu=1,\alpha=2$ and $\nu=0.8,\alpha=1.6$ lead to the same scaling $t^{1/2}$, although in the case $\nu=0.8,\alpha=1.6$, the process $X(t)=\tilde{X}(S_t)$ is non-Markov, in contrast to a standard Gaussian diffusion obtained for $\nu=1, \alpha=2$. Thus, the competition between subdiffusion and L\'evy flights questions standard ways of discrimination between normal (Markov, $\langle [x-x(0)]^2 \rangle \propto t$) and anomalous (generally, non-Markov $\langle [x-x(0)]^2 \rangle \propto t^\delta$) diffusion processes.

Indeed, for $\nu=1$, the process $X(t)$ is not only $1/\alpha$ self-similar but it is also memoryless (i.e. Markovian). In such a case, the asymptotic PDF $p(x,t)$ is $\alpha$-stable \cite{janicki1994,weron1995,weron1996,dybiec2006} with the scale parameter $\sigma$ growing with time like $t^{1/\alpha}$, cf. Eq.~(\ref{eq:charakt1}). This is no longer true for subordination with $\nu<1$ when the underlying process becomes non-Markovian and the spread of the distribution follows the $t^{\nu/\alpha}$-scaling (cf. Fig.~\ref{fig:quantiles}, right panels).

Some additional care should be taken when discussing the scaling character of moments of $p(x,t)$ \cite{bouchaud1990,fogedby1998,jespersen1999}. Clearly, L\'evy distribution (with $\alpha<2$) of jump lengths leads to infinite second moment (see Eqs.~(\ref{eq:charakt1}) and (\ref{eq:charakt2}))
\begin{equation}
\langle x^2 \rangle = \int_{-\infty}^{\infty} x^2 l_{\alpha,\beta}(x;\sigma) dx = \infty,
\end{equation}
irrespectively of time $t$.
Moreover, the mean value $\langle x \rangle$ of stable variables is finite for $\alpha >  1$ only ($\langle x \rangle=0$ for symmetric case under investigation). Those observations seem to contradict demonstration of the scaling visible in Fig.~\ref{fig:stdev} where standard deviations derived from ensembles of trajectories are finite and grow in time according to a power law.
A nice explanation of this behavior can be given following argumentation of Bouchaud and Georges \cite{bouchaud1990}: Every finite but otherwise arbitrarily long trajectory of a L\'evy flight, i.e. the stochastic process underlying Eq.~(\ref{eq:ffpe}) with $\nu=1$, is a sum of finite number of independent stable random variables. Among all summed $N$ stable random variables there is the largest one, let say $l_c(N)$. The asymptotic form of a stable densities
\begin{equation}
l_{\alpha,\beta}(x;\sigma) \propto x^{-(1+\alpha)},
\label{eq:asymptotics}
\end{equation}
together with the estimate for $l_c(N)$ allow one to estimate how standard deviations grows with a number of jumps $N$. In fact, the largest value $l_c(N)$ can be estimated from the condition
\begin{equation}
N\int_{l_c(N)}^{\infty}l_{\alpha,\beta}(x)dx\approx 1.
\label{eq:trials}
\end{equation}
which locates most of the ``probability mass'' in events not exceeding the step length $l_c$ (otherwise, the relation states that $l_c(N)$ occurred at most once in $N$ trials, \cite{bouchaud1990}). Alternatively, $l_c(N)$ can be estimated as a value which maximizes probability that the largest number chosen in $N$ trials is $l_c$
\begin{equation}
l_{\alpha,\beta}(l_c)\left[ \int\limits_0^{l_c} l_{\alpha,\beta}(x)dx \right]^{N-1}=l_{\alpha,\beta}(l_c)\left[ 1 - \int\limits_{l_c}^\infty l_{\alpha,\beta}(x)dx \right]^{N-1}.
\label{eq:minimalization}
\end{equation}
By use of Eqs.~(\ref{eq:trials}) and (\ref{eq:asymptotics}), simple integration leads to
\begin{equation}
l_c(N)\propto N^{1/\alpha}.
\label{eq:threshold}
\end{equation}
Due to finite, but otherwise arbitrarily large number of trials $N$, the effective distributions becomes restricted to the finite domain which size is controlled by Eq.~(\ref{eq:threshold}). Using the estimated threshold, see Eq.~(\ref{eq:threshold}) and asymptotic form of stable densities, see Eq.~(\ref{eq:asymptotics}), it is possible to derive an estimate of $\langle x^2 \rangle$
\begin{equation}
\langle x^2 \rangle \approx \int^{l_c}x^2 l_{\alpha,\beta}(x) dx\approx \left( N^{1/\alpha} \right)^{2-\alpha}=N^{2/\alpha-1}.
\end{equation}
Finally, after $N$ jumps
\begin{equation}
\langle x^2 \rangle_N=N\langle x^2 \rangle \propto N^{2/\alpha}.
\label{eq:variancescaling}
\end{equation}
Consequently, for L\'evy flights standard deviation grows like a power law with the number of jumps $N$.
In our CTRW scenario incorporating competition between long rests and long jumps, the number of jumps $N=N(t)$ grows sublinearly in time, $N\propto t^{\nu}$, leading effectively to $\langle x^2 \rangle_N\propto t^{2\nu/\alpha}$ with $0<\nu<1$ and $0<\alpha<2$. Since in any experimental realization tails of the L\'evy distributions are almost inaccessible and therefore effectively truncated, analyzed sample trajectories follow the pattern of the $t^{\nu/\alpha}$ scaling, which is well documented in Fig.~\ref{fig:stdev}.

%
%
\section{Discriminating memory effects}

Clearly, by construction, for $\nu < 1$ the limiting ``anomalous diffusion'' process $X(t)$ is non-Markov. This feature is however non-transparent when discussing statistical properties of the process by analyzing ensemble-averaged mean-square displacement for the parameters set $\nu/\alpha=1/2$, (e.g. $\nu=0.8,\alpha=1.6$) when contrary to what might be expected, $\langle [x-x(0)]^2 \rangle \propto t$, similarly to the standard, Markov, Gaussian case.
This observation implies a different problem to be brought about: Given an experimental set of data, say time series representative for a given process, how can one interpret its statistical properties and conclude about anomalous (subdiffusive) character of underlying kinetics? The similar question has been carried out in a series of papers discussing use of transport coefficients in systems exhibiting weak ergodicity breaking (see \cite{he2008} and references therein).

To further elucidate the nature of simulated data sets for $\nu/\alpha=1/2$, we have adhered to and tested formal criteria \cite{vankampen1981,fulinski1998,dybiec2009b,dybiec-sparre} defining the Markov process. The standard formalism of space- and time- continuous Markov processes requires fulfillment of the Chapman-Kolmogorov equation ($t_1>t_2>t_3$)
\begin{equation}
 P(x_1,t_1|x_3,t_3) = \sum_{x_2}P(x_1,t_1|x_2,t_2)P(x_2,t_2|x_3,t_3).
\label{eq:ch-k}
\end{equation}
along with the constraint for conditional probabilities which for the ``memoryless'' process should not depend on its history. In particular, for a hierarchy of times $t_1>t_2>t_3$, the following relation has to be satisfied
\begin{equation}
 P(x_1,t_1|x_2,t_2)=P(x_1,t_1|x_2,t_2,x_3,t_3).
\label{eq:cond}
\end{equation}
Eqs.~(\ref{eq:ch-k}) and (\ref{eq:cond}) have been used to directly verify whether the process under consideration is of the Markovian or non-Markovian type. From Eq.~(\ref{eq:ch-k}) squared cumulative deviation $Q^2$ between LHS and RHS of the Chapman-Kolmogorov relation summed over initial ($x_3$) and final ($x_1$) states has been calculated \cite{fulinski1998}
\begin{eqnarray}
 Q^2 & = & \sum\limits_{x_1,x_3}\bigg[P(x_1,t_1|x_3,t_3) \nonumber \\
 & & \qquad\quad - \sum\limits_{x_2}P(x_1,t_1|x_2,t_2)P(x_2,t_2|x_3,t_3)\bigg]^2.
\label{eq:ch-k-averaged}
\end{eqnarray}
The same procedure can be applied to Eq.~(\ref{eq:cond}) leading to
\begin{eqnarray}
M^2 & = & \sum_{x_1,x_2,x_3}\Big[P(x_1,t_1|x_2,t_2) \nonumber \\
& & \qquad\quad - P(x_1,t_1|x_2,t_2,x_3,t_3)\Big]^2.
\label{eq:cond-averaged}
\end{eqnarray}

\begin{figure}[!ht]
\begin{center}
\includegraphics[angle=0,width=8.0cm]{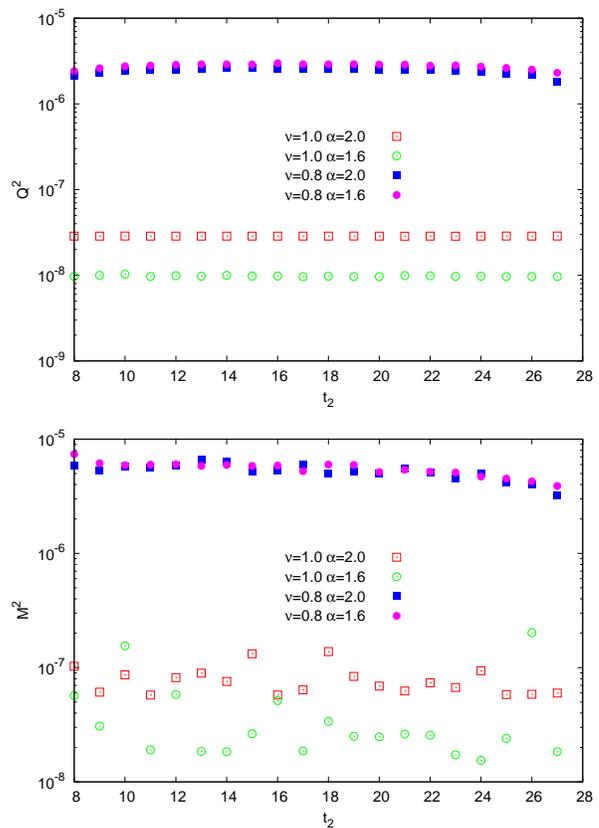}
\caption{(Color online) Squared sum of deviations $Q^2$, see Eq.~(\ref{eq:ch-k-averaged}), (top panel) and $M^2$, see Eq.~(\ref{eq:cond-averaged}), (bottom panel) for $t_1=27$, $t_3=6$ as a function of the intermediate time $t_2$. 2D histograms were created on the $[-10,10]^2$ domain. 3D histograms were created on the $[-10,10]^3$ domain. Due to non-stationary character of the studied process the analysis is performed for the series of increments $x(t+1)-x(t)$. Simulation parameters as in Fig.~\ref{fig:trajectories}.}
\label{fig:ch-k}
\end{center}
\end{figure}

Figure~\ref{fig:ch-k} presents evaluation of $Q^2$ (top panel) and $M^2$ (bottom panel) for $t_1=27$ and $t_3=6$ as a function of the intermediate time $t_2=\{7,8,9,\dots,25,26\}$. It is seen that deviations from the Chapman-Kolmogorov identity are well registered for processes with long rests when subdiffusion wins competition with L\'evy flights at the level of sample paths.
The tests based on $Q^2$ (see Eq.~(\ref{eq:ch-k-averaged})) and $M^2$ (see Eq.~(\ref{eq:cond-averaged})) have comparative character. The deviations $Q^2$ and $M^2$ are about three order of magnitudes higher for the parameter sets $\nu=0.8, \alpha=2.0$ and $\nu=0.8, \alpha=1.6$ than $Q^2$ and $M^2$ values for the Markovian counterparts with $\nu=1$ and $\alpha=2$, $\alpha=1.6$, respectively. Performed analysis clearly demonstrates non-Markovian character of the limiting diffusion process for $\nu<1$ and the findings indicate that scaling of PDF, $p(x,t)=t^{-1/2}p(xt^{-1/2},1)$ or, in consequence, scaling of the variance
$\langle [x-x(0)]^2 \rangle \propto t$ and interquantile distances (see Fig.~\ref{fig:quantiles}) do not discriminate satisfactory between ``normal'' and ``anomalous'' diffusive motions \cite{zumofen1995}. In fact, linear in time spread of the second moment does not necessarily characterize normal diffusion process. Instead, it can be an outcome of a special interplay between subdiffusion and L\'evy flights combined in the subordination $X(t)=\tilde{X}(S_t)$. The competition between both processes is better displayed in analyzed sample trajectories $X(t)$ where combination of long jumps and long trapping times can be detected, see Fig.~\ref{fig:trajectories}.

%
%
\section{Conclusions}

Summarizing, by implementing Monte Carlo simulations which allow visualization of stochastic trajectories subjected to subdiffusion (via time-subordination) and superdiffusive L\'evy flights (via extremely long jumps in space), we have demonstrated that the standard measure used to
discriminate between anomalous and normal behavior cannot be applied
straightforwardly. The mean square displacement alone, as derived from the (finite) set of time-series data does not provide full
information about the underlying system dynamics. In order
to get proper insight into the character of the motion, it is necessary to perform analysis of individual trajectories.
Subordination which describes a transformation between a physical time and an operational time of the system \cite{sokolov2000,magdziarz2007b} is responsible for unusual statistical properties of waiting times between subsequent steps of the motion. In turn,
L\'evy flights are registered in instantaneous long jumps performed by a walker. Super- or sub- linear character of the motion in physical time is dictated by a coarse-graining procedure, in which fractional time derivative with the index $\nu$ combines with a fractional spatial derivative with the index $\alpha$.
Such situations may occur in motion on random potential surfaces where the presence of vacancies and other defects introduces both --
spatial and temporal disorder \cite{sancho2004}.
We believe that the issue of the interplay of super- and sub-diffusion with a crossover resulting in a pseudo-normal paradoxical diffusion may be of special interest in the context of e.g. facilitated target location of proteins on folding heteropolymers \cite{belik2007} or in analysis of single particle tracking experiments \cite{lubelski2008,he2008,metzler2009,bronstein2009}, where the hidden subdiffusion process can be masked and appear as a normal diffusion.

%
\begin{acknowledgments}
This project has been supported by the Marie Curie TOK COCOS grant (6th EU Framework
Program under Contract No. MTKD-CT-2004-517186) and (BD) by the Foundation for Polish Science.
The authors acknowledge many fruitfull discussions with Andrzej Fuli\'nski, Marcin Magdziarz and Aleksander Weron.
\end{acknowledgments}

%
%

\end{document}